\documentclass[letterpaper, 10pt, conference]{ieeeconf} 

\IEEEoverridecommandlockouts                              
\usepackage{graphics} 
\usepackage{epsfig} 
\usepackage{mathptmx} 
\usepackage{times} 
\usepackage{amsmath} 
\usepackage{amssymb}  

\begin{document}
\title{\LARGE \bf Nonlinear Instabilities of D2TCP-II}
\author{Abhisek Mukhopadhyay$^{1}$ \thanks{$^{1}$A.Mukhopadhyay is with the department of Electronics and Communication Engineering, National Institute of Technology, Durgapur, West Bengal, India-71320.%
 {\small E-Mail- abhisekmukhopadhyay[at]gmail. com}} \and Priya Ranjan$^{2}$%
\thanks{$^{2}$ P. Ranjan is with the department of Electronics and Communication Engineering, Templecity Institute of Technology and Engineering, Khurda, Orissa, India-752057.  {\small E-mail-pranjan[at]gmail. com} }}%

\maketitle
\thispagestyle{empty}
\pagestyle{empty}
\begin{abstract}
In the era of heavy-duty transmission control protocols (TCP), adapted for extremely hi-bandwidth data-centers; the fundamental question of stable interaction with either proposed/customized active queue management(AQM) or popularly discussed Random Early Detection (RED) remains a hotly debated issue.  While there are claims of ``oscillation'' only dynamical behavior, there are equally large number of claims which demonstrate the chaotic nature of different flavors of TCP and their AQM interaction.  In this work, we provide a sound and analytical mathematical model of DTCP/D2TCP and study their interaction with threshold based packet marking policy.  Our work shows that for a simple scenario this interaction is chaotic in nature and has large variability in dynamical behavior over orders of magnitude changes in parameter range as demonstrated by bifurcation diagrams.    We conclude with numerical simulation evidence that chaotic behavior of protocols is inherent in their design which they inherit from their early vanilla TCP days, and it has serious implications for data-center throughput, load batching and collapse in Incast kind of scenario. 
 \end {abstract}
\section{Introduction} 
In the era of record breaking memory to memory throughput, dynamical behavior of Internet protocols has become key to achieving predictability, optimal utilization of resources and dynamic stability in throughput allocation among participating agents/devices/entities. With the recent  proposals of DCTCP, D2TCP, CTCP, D3 etc, a new research area of super-optimal and dynamically oscillating behavior of transport protocols has been forging itself. 

We based our analysis and discrete time models in contrast to the early claims of periodic only behavior in DCTCP and its different relatives and derivatives. We show again that discrete time models are the right framework to think about next generation successors of TCP and chaos remains a dominant phenomenon in the case of wildly fluctuating parameter ranges. In particular, we compare its dynamical behavior with switched power electronic circuits and impact oscillators. 
\section{Objective of the study}
The objectives of our study are as follows:\\
\begin{enumerate}
\item Build a discrete time mathematical model of DCTCP-AQM interaction. 
\item Use techniques from discrete time maps to analyze their  behavior. 
\item Simulate and demonstrate chaotic behavior. 
\end{enumerate}
\section{ DCTCP congestion control algorithm:An overview}
In this section we describe qualitatively the DCTCP algorithm\cite{c1} with its major algorithmic details so as to lay a foundation for our discrete modeling. We also describe the fact that D2TCP \cite{c2} can be built upon DCTCP through certain specific changes and we also argue that D2TCP is more general and that DCTCP forms a special case of D2TCP. 
\subsection{\normalsize A simple active queue management scheme at the Switch:}The marking scheme involves the use of only a single parameter $ K$.  An arriving packet is marked with its CE (Congestion Experienced)  bit set to 1 iff upon its arrival the instantaneous queue size is greater than $K$. 
\subsection{\normalsize ECN Echo at receiver} A difference between conventional TCP receivers and DCTCP receiver is in the way the status of the incoming CE bits are conveyed back to the sender.  For the DCTCP receiver ACKs (ACK- Acknowledgment Packet) every single incoming packet and  its ECN(External Congestion Notification)- Echo flag bit set to 1 when a marked packet  (CE=1) is received.  DCTCP tries to convey the exact sequence of the received code points.  This is accomplished by sending a delayed ACK (1 cumulative ACK for a set of m received packets) which contains the sequence of ECN- Echo flag based on last m incoming packets.  Thus the controller at the sender can make out the congestion status in the queue upon reception of the delayed ACK. 
\subsection{\normalsize Controller at the sender}The sender maintains a running estimate of the fraction of packets that are marked. Called $ \alpha$, Which is updated in every$ RTT$ (Round Trip Time) as follows:
\begin{equation}
\alpha \longleftarrow (1-g)\cdot\alpha +g\cdot F \label{alph}
\end{equation}
$g$ here is the weight associated to the marked fraction. 
$F$  is the fraction of packets marked and is given by the drop probability in the current window.  
Given the fact that the sender receives Marks for every packet sent when the queue length is higher than the threshold and does not receive any marks when the queue length is below threshold $K$  thus \eqref{alph}  implies that  $\alpha$ estimates the probability that the queue size is greater than $K$. Essentially, $\alpha$ close to 0 implies low levels of congestion and when close to 1 implies high levels of congestion. 
\subsection{\normalsize Window size control at the sender :}The only difference between a DCTCP Sender and a TCP Sender is how each react on receiving the ACK with ECN flag on. In the slow start phase the Window size increases slowly in both the cases to estimate the available bandwidth.  On receiving congestion notification The TCP always controls congestion by halving the window size and then slowly increasing additively in the congestion avoidance phase.  
Where as in the DCTCP scheme when it senses congestion the window size is modified as:\newline
\begin{equation}
cwnd \longleftarrow (1- \frac{\alpha}{2})\cdot cwnd \label{wind}
\end{equation}
Thus the congestion extent determines the change in the window size.  It is only slightly reduced at low levels of congestion and when congestion levels are high ,say 1 then the congestion window is reduced by half as in TCP. 
\emph{This is the chief contribution of DCTCP in maintaining a low queue length without compromising throughput}.

     In Deadline Aware Data Center TCP (D2TCP) all the essential components of DCTCP is kept same but the $\alpha$ (congestion history parameter)  is modified as :\\
$\alpha=\alpha^\gamma$
Where $\gamma$  is the deadline factor. A high $\gamma$  implies a closer deadline and a low value of $\gamma$  implies far deadline.  So DCTCP is a special case of D2TCP with $\gamma = 1$  which implies long lived flows with no deadline. 
\section{Conventional RED Controller at Switch}
The RED \cite{c3} mechanism is available as an External Congestion Notification (ECN) mechanism in modern switches.  This section gives a brief discussion on the RED (Random Early Detection) at the switch and the modification that needs to be made in order to support the DCTCP queue management scheme without the need of extra external hardware. 
 We present here a discrete model of the RED system and the sampling interval is one RTT. 

The RED module calculates exponentially weighted moving average of the queue size at the switch end. 

 Let $ w$  be the exponential averaging weight and $q_k$  be the instantaneous queue size. 
 On each packet arrival the RED algorithm updates its average queue length $ (\overline {q_k}) $ to: \newline
\begin{equation}
 \overline {q}_{k+1}=(1-w)\cdot \overline {q}_{k}+w\cdot q_{k+1}
\end{equation}

If the average queue length is below a certain advertised threshold $q_{min}$ then the packet is admitted into the queue for subsequent transmission through the link and when it exceeds $q_{max}$  it is either dropped or marked as per requirement. If it lies in the range of $qmin$ and $qmax$ then it is dropped/marked with a probability $p$  given by:
\begin{equation}
p_{k+1}=\begin{cases}
0 &\text{if}{} \quad  \overline{q}_{k+1}<q_{min}\\
\frac{\overline{q}_{k+1}-q_{min}}{q_{max}-q_{min}}\cdot p_{max} & \text{if}\quad {}  q_{min}\leq \overline{q}_{k+1}\leq q_{max}\\
1 &\text{if}\quad \overline{q}_{k+1}>q_{max}
\end{cases}\label{red}
\end{equation}
\subsection{\normalsize Implementation of DCTCP packet marking strategy through RED}
To implement DCTCP packet marking strategy through RED we need to:
\begin{enumerate}
\item Set $q_{min}=q_{max}=K$ 
\item The decision making should be with respect to the instantaneous queue size $(q_{k})$ instead of the average queue size. 
\end{enumerate}
Thus implementing the above \eqref{red}  becomes
\begin{equation}
p_{k+1}=\begin{cases}
0 &\text{if}{} \quad  q_{k+1}\leq K\\

1 &\text{if}\quad q_{k+1}>K
\end{cases}\label{dctcpred}
\end{equation}
Thus,\eqref{dctcpred} gives us the hard control required in the case of DCTCP algorithm. 

\section{ Derivation of  a discrete model for a network employing DCTCP at the sender end and modified RED at the Switch. }
\subsection{\normalsize Background}
We model the D2TCP system along with its implicit congestion control mechanism at the switch implemented through available RED module, as a discrete time map obtained by periodically sampling the system state.  Since window size and queue size behave as step functions of an RTT,one RTT is the sampling interval that captures their changes.  Based on the DCTCP algorithm we propose our discrete model that takes window size and congestion history as state variables.  We develop the expressions for a single active connection so as to present the dynamics of the queue size and study the effect of the changes of parameter on the it.  The model can be extended without the loss of generality to $N$ synchronized connections. 
\subsection{\normalsize The model}
A complex network employing DCTCP is essentially a feedback loop in which the senders adjust their rate by observing the rate of packet loss i. e by  a feedback from the router. 
\begin{figure}[htb]
\begin{center}
\includegraphics[height=1.5in,width=2.5 in]{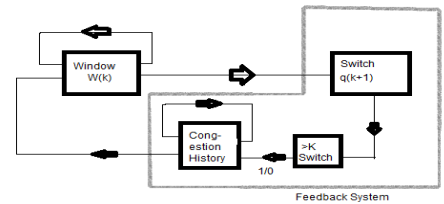}
\caption{Figure showing the feedback mechanism of DCTCP}\label{fig}
\end{center}
\end{figure}
Each flow at a router sends packets that are queued in the shallow buffer\footnote{Shallow Buffer means buffers having low capacity in high speed switches} for subsequent transmission through the link. The packets are marked following a control strategy employed at the router end. DCTCP employs a \textsl  {bursty} marking strategy at the router following \eqref{dctcpred}. When the sender notices the packets are being marked it adjusts its sending rate accordingly,In the case of DCTCP the window size is reduced following a congestion history parameter $\alpha$. The logical structure of DCTCP congestion controller is given in Figure 1. This forms the  basis of formulation of the discrete mathematics in subsequent sections and its physical implications. A similar analysis has been done in\cite{c4} to model the behavior of TCP-RED interaction. 
\subsection{\normalsize A discrete modeling of the system}
\subsubsection{ Assumptions}
We make the following assumptions in our model. 
\begin{itemize}
\item There is one active connection and the connection is long lived that is, there is always sufficient data to
send. 
\item The ACK packets are never lost. 
\item Round Trip propagation delay in the link and the packet size is constant. 
\item The state variables of the system are sampled at the end of every RTT interval. 
\item The process starts with an empty buffer. 
\end{itemize}
\subsubsection{Dynamical Modeling of the parameters}
We first model the dynamics of the window size $W$ using a discrete approximation of \eqref{wind} for more general D2TCP case.  Let $W_k$ ,$ q_k$ ,$\alpha_k$ and $\gamma$  be the window size ,instantaneous queue size ,the weighted fraction of the
marked packets and the deadline of the flow at the end of the $ k^{th}$ sampling interval.  The window size for the next interval is determined by the queue size in that interval \footnote{The window size calculated at the $k^{th}$ interval determines how many packets to schedule for transfer in the next interval so the window size in the $k^{th}$ interval participates in the queue generation in the ${k+1}^{th}$ interval}, it is given by:
\begin{equation}
W_{k+1}=\begin{cases}
(1-\frac{\alpha_k^\gamma}{2})\cdot W_k  &\text{if}\quad  q_{k+1}>K,\\
 W_k +1 &\text{otherwise. }
\end{cases}\label{window}
\end{equation}

The queue size at the sampling period $k+1$ can be determined from the window size at $k$ and the
queue size during $k^{th}$ interval as:
\begin{equation}
\begin{aligned}
q_{k+1}&=q_k+W_k-\frac{C}{M} \cdot RTT_{k+1}\\
&=q_k+W_k-\frac{C}{M} \cdot (q_k\cdot \frac{M}{C}+d)\\
&=W_k-\frac{C\cdot d}{M}\\
\end{aligned}
\label{queue}
\end{equation}
\begin {description}
\item{In \eqref{queue} ,}
$\frac{C}{M}$ gives the link capacity in packets/s. \\
$RTT_{ k+1}=d+q_k\cdot\frac{M}{C}$; gives the Round Trip Time at the ${k+1}^{th}$ interval\\
$ d$- Round trip propagation delay in the link\\
 $C $-Capacity and $ M$ -Packet Size. 
\end{description}
As the buffer is limited by its size $B$. And a negative queue size has no meaning,we can set realistic bounds on $q_{k+1}$  as,
\begin{equation}
q_{k+1}=min(max(W_k-\frac{C\cdot d}{M},0),B)
\end{equation}
The probabilistic \emph{bursty}  marking scheme at the switch remains \eqref{dctcpred}. \\
The congestion history parameter is updated every $RTT$ as,
\begin{equation}
\alpha_{k+1}=(1-g)\cdot \alpha_k +g\cdot p_{k+1}
\label{alph}
\end{equation}
Or alternatively,using \eqref{dctcpred},
\begin{equation}
\alpha_{k+1}=\begin{cases}
(1-g)\cdot \alpha_k +g &\text{if}\quad q_{k+1}>K,\\
(1-g)\cdot \alpha_k &\text{otherwise}. 
\end{cases}
 \label{conges}
\end{equation}
\subsection{\normalsize The Map}
In this model as the instantaneous queue size is taken into consideration for the marking of packets so, the queue build up has no memory and depends only on the window size of the previous instant. Thus, the queue size $q_{k}$ is not an independent parameter. The independent parameters are the congestion history ($\alpha_k$) and the window size ($W_k$). \\Substituting \eqref{queue} in \eqref{window} we have,
\begin{equation}
W_{k+1}=\begin{cases}
(1-\frac{\alpha_k^\gamma}{2})\cdot W_k  &\text{if}\quad  W_k-\frac{C\cdot d}{M}>K,\\
 W_k +1 &\text{otherwise. }
\end{cases}\label{map1}
\end{equation}
and substituting  \eqref{queue} in \eqref{conges} we have,
\begin{equation}
\alpha_{k+1}=\begin{cases}
(1-g)\cdot \alpha_k +g &\text{if}\quad  W_k-\frac{C\cdot d}{M}>K,\\
(1-g)\cdot \alpha_k &\text{otherwise}. 
\end{cases}\label{map2}
\end{equation}
thus \eqref{queue},\eqref{map1},\eqref{map2} give the map form that can be used to model the dynamics of the system. 
\subsection{Mathematical analysis of phase space and the borders. }
The map of $W$ has two distinct regions of operation separated by a border $q_{k+1}=K$. This border on the instantaneous queue size reflects on the sender side as a new border: $W_k=K^{*}$ where $K^{*}=\frac{C\cdot d}{M}+K$. In The additive increase phase where if the $W_k$ is less than a threshold value $K^{*}$ implying, $q_{k+1}<K$  the definition of the map in \eqref{map1} says that the window size slowly increases and this additive increase is independant of $\alpha$.  As a result of the additive increase, the window size after a certain time hits the threshold ($K^{*}$) and  the map becomes discontinuous as the control mechanism kicks in, making the evolution of the window dependent on both the state variables ($\alpha_k \& W_k$). The use of congestion history to determine the next window size when congestion is detected is the key feature of DCTCP,thus a high $\alpha_k$ implies a higher cut in the window size in the next stage and vice versa. These features of the evolution of the window size are demonstrated graphically in Fig. 2 \& Fig. 3. 

The evolution of $\alpha$ also follows a similar algorithm with borders playing the role for the discontinuity induced. The Congestion history for the next step is determined using the knowledge of the queue size in the current step. As long as the system senses congestion i. e $q_{k+1}>K$ the congestion history successively increases, simultaneously window size multiplicatively decreases until it hits the border $K^{*}$ from the right side.  When the $W_k$ becomes less than $K^{*}$ i. e there is \emph{'no congestion' feedback},$\alpha$ successively decreases and correspondingly  window size goes into the additive increase phase until congestion arises again,i. e this time the congestion window hits the border $K^{*}$ from the left. 

Thus, the congestion control system operates by successively switching between two subsystems as a result of the system states colliding with border values.  This in turn leads to chaotic dynamics through \emph{border collision bifurcation} as we will demonstrate later. 

The first and the second return maps for $\alpha$ are given in Fig. 4 for the evolution of $q_{k+1}=35$ through successive iterations. 
for generating the maps we use,
\quad $\gamma=1$,$d=30\mu s$,$C=10 Gbps$,$g=1/16$,$K=15 packets$,   $M=1 Kb$

   \begin{figure}[tphb]

\includegraphics[height=4cm, width=6cm]{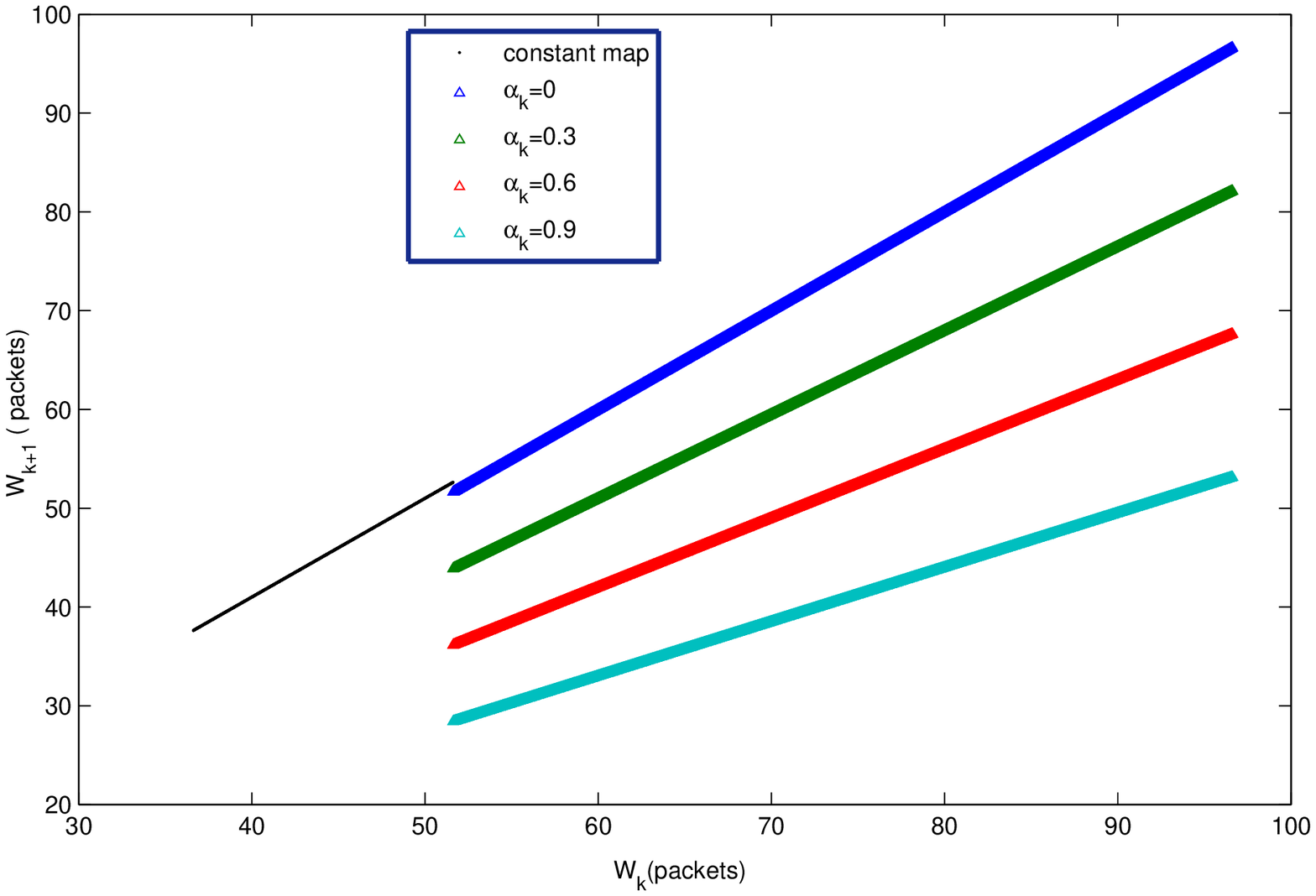}
      \caption {\scriptsize  Figure showing the $ 1^{st}$ return  map of $W$ at different congestion levels when K=15}
      \includegraphics[height=4cm, width=6cm]{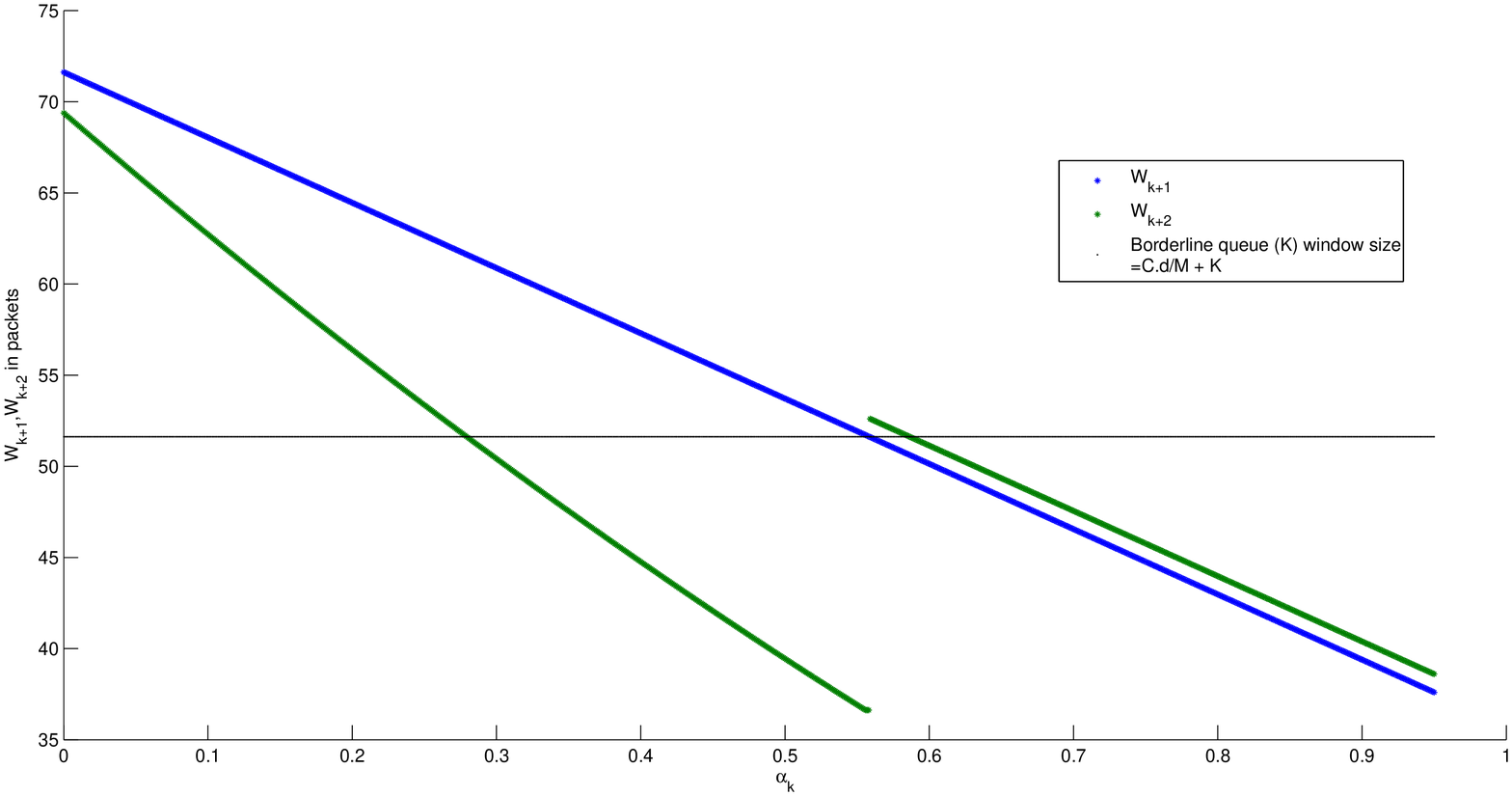}
      \caption {\scriptsize  Figure showing the $ 1^{st}$ and $2^{nd}$ return  map of $W_k=71. 62 (q_{k+1}=35)$ at different congestion levels when K=15}
     \includegraphics[height=4cm ,width=6cm]{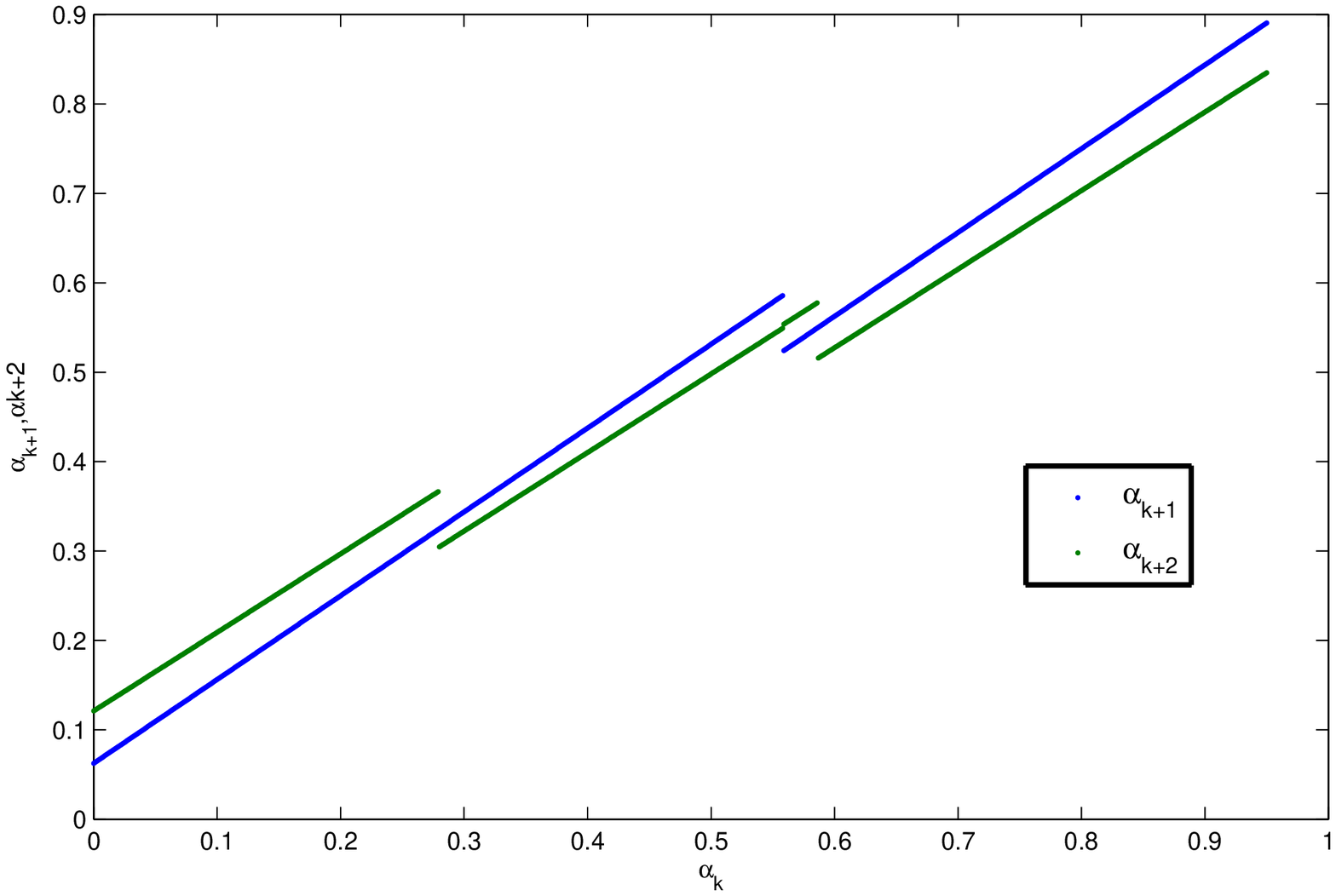}
      \caption {\scriptsize  Figure showing the $ 1^{st}$  and $2^{nd}$ return  map of $\alpha$ when $W_k=71. 62 (q_{k+1}=35)$ when K=15}
      \label{maps}
   \end{figure}
\section {Numerical Analysis}
The behavior of the map can be explored in parameter space numerically,in order to find interesting dynamical phenomena. 

A whole range of dynamical scenarios are presented in this section. The effect of the changes on the system described by  \emph{piecewise smooth discontinuous maps} are reflected through \emph{bifurcation diagrams}. 
\subsection{\normalsize Bifurcation diagrams}
When the future state of a dynamical variable is dependent on a particular parameter(s) then variation in the parameter(s) would result in change in the dynamical behavior of the system on a whole. A bifurcation diagram shows the qualitative changes in the nature or the number of steady state solutions of the dynamical system as a parameter varies. On the horizontal axis we plot the parameter value. The vertical axis displays a measure of the corresponding fixed points or periodic orbits\footnote{Mathematically speaking, topological invariants},which coincides with the queue build up in the present context. 

The way to read a bifurcation diagram is to fix a point in the horizontal axis and draw a vertical line through it. The number of points where the bifurcation curve intersects the vertical line gives us the number of equilibrium points in that given configuration. A single point implies there is a stable fixed point and multiple points imply a periodic long-term behavior of the system. 

We investigate the dynamical behavior of the instantaneous queue size in a DCTCP (D2TCP) system with a single connection when the parameters like Marked fraction weight ($g$),the round trip propagation delay ($d$),deadline($\gamma$) and the marking threshold ($K$) are varied. We draw the bifurcation diagram by plotting the successive maxima and minima of the instantaneous queue size\footnote{Transient responses are eliminated by long time evaluation of the system states. Steady state solutions ought to be used}(in packets)  against a  parameter setup. 
Other system parameters are kept fixed at:

$C=10Gbps ; M=1 Kb ; B=200 packets$. 
\subsection{\normalsize Bifurcation parameter : Marked fraction weight ($g$)}
We present the bifurcation diagram  with $g$ as the bifurcation parameter which varies from 0. 001 to 0. 1 through a step size of 0. 0001 ,and other parameters are  given as. \\
Case 1:\\
When $\gamma=1$ ; $d= 1 ns$ ; K=20 in Fig. 5\\
Case 2:\\
When $\gamma=1$ ; $d= 100\mu s$ ; K=20 in Fig. 6
 \begin{figure}[tphb]

\includegraphics[height=4.5cm, width=7cm]{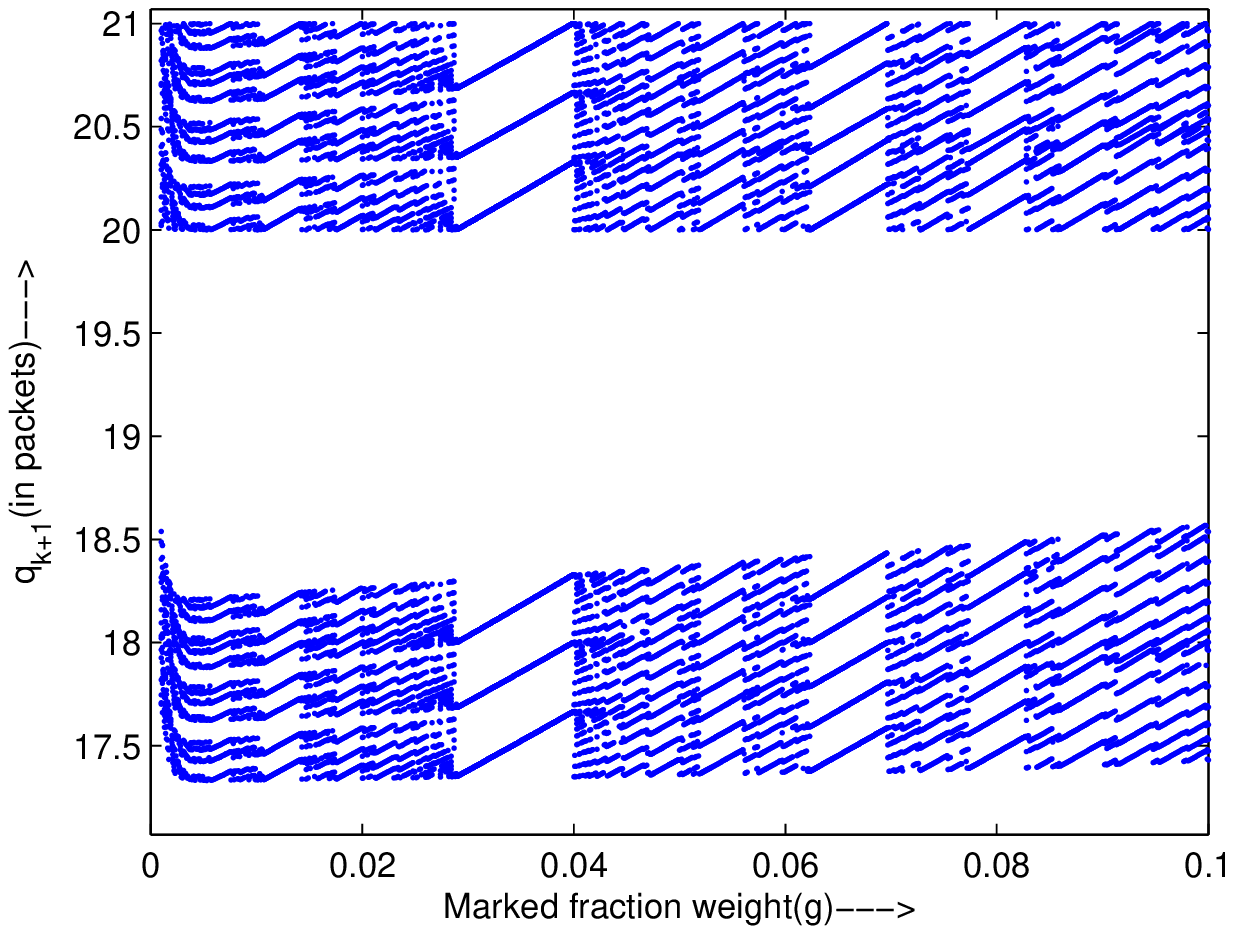}
      \caption {\scriptsize  The bifurcation diagram with $g$ as the bifurcation parameter and $d=1 ns$}
\includegraphics[height=4.5cm, width=7cm]{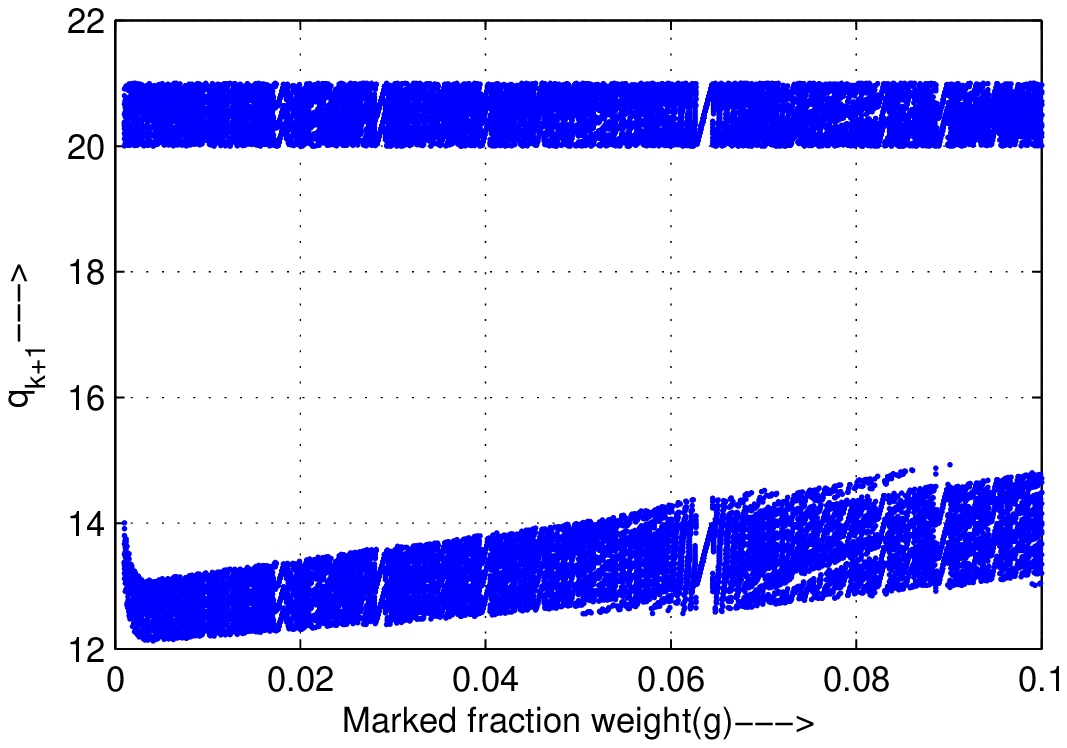}
      \caption {\scriptsize  The bifurcation diagram with $g$ as the bifurcation parameter and $d=100 \mu s$}
\end{figure}

In the bifurcation diagram for parameter setup in case 1. As the Round Trip Delay is very small so the packet marks are fed-back faster so we should expect stable dynamical behaviour. The bifurcation diagram of the system has a wide range of $g$ for which stable periodic solutions exist which is shown by the region of the diagram which has a finite number of fixed points. Like the region having 6 fixed points correspond to a stable period-3 orbit in the time series plot of the queue size. But for a range of $g$ we have bounded but locally unstable range of fixed points, that generates a band like structure in the diagram for the corresponding ranges. 

In case 2. The delay is made larger so the feedback system is slow. This gives rises to more complicated dynamics. The periodic windows disappear, the amplitude of oscillations increase and ranges of $g$ for chaotic oscillation increases.  This complication in behaviour as the delay is increased is consistent with the results in \cite{p1}\cite{p2}. 
 \subsection{\normalsize Bifurcation parameter : Round trip delay ($d$)}
The Round trip delay(d) is the parameter which varies quite a lot in practical networking scenarios. We draw the bifurcation diagram (Fig. 7) for delay varying from the range of $ 1 ns $ to $ 100\mu s$. Other parameters set up as: \\
$K=20;\gamma=1;g=0.037. $\\
\begin{figure}[tphb]
\centering
\includegraphics[height=4.5cm, width=8cm]{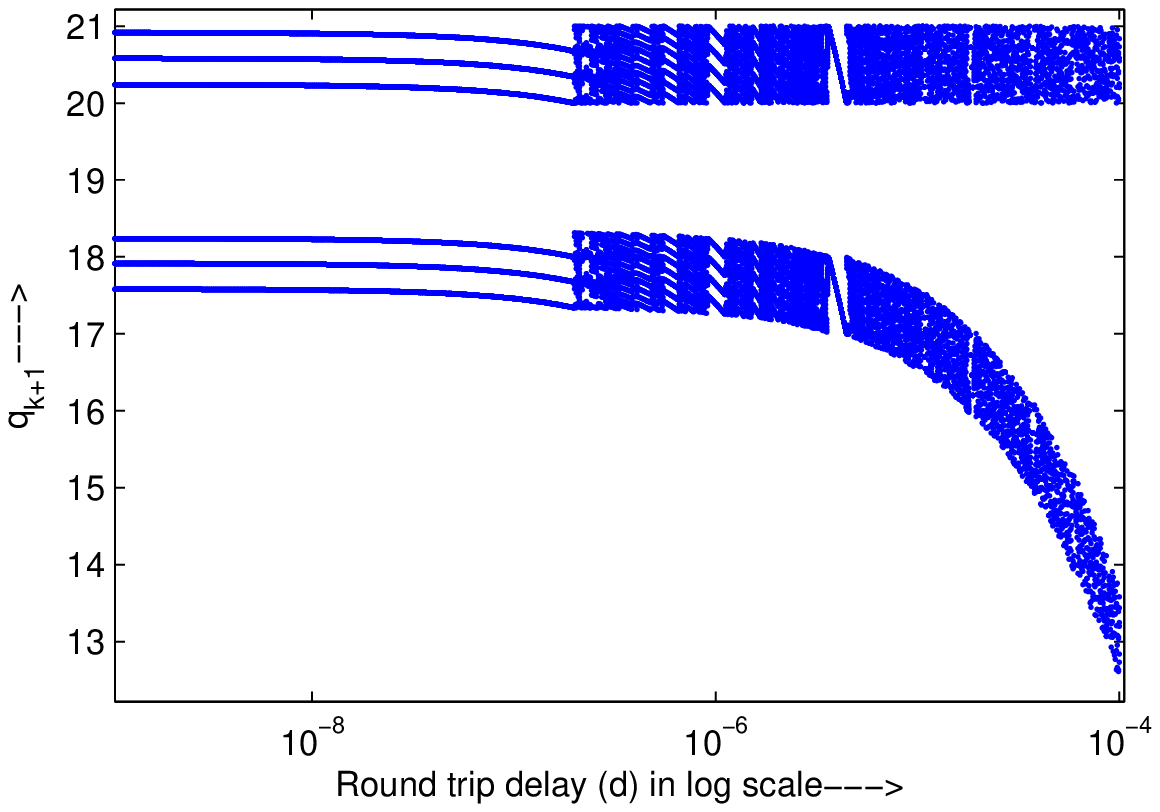}
      \caption {\scriptsize  The bifurcation diagram with $d(in seconds)$ as the bifurcation parameter}
\includegraphics[scale=0.6]{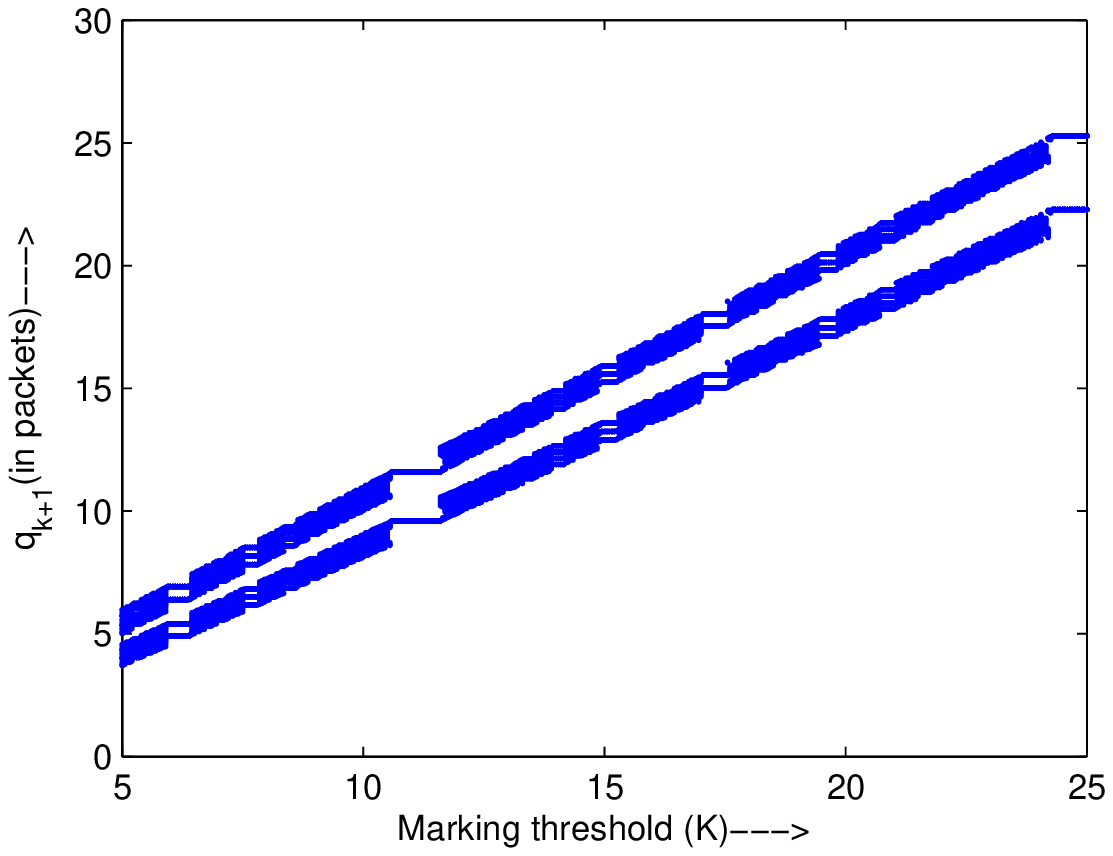}
      \caption {\scriptsize  The bifurcation diagram with $K$ as the bifurcation parameter}
\includegraphics[height=4.5cm, width=8cm]{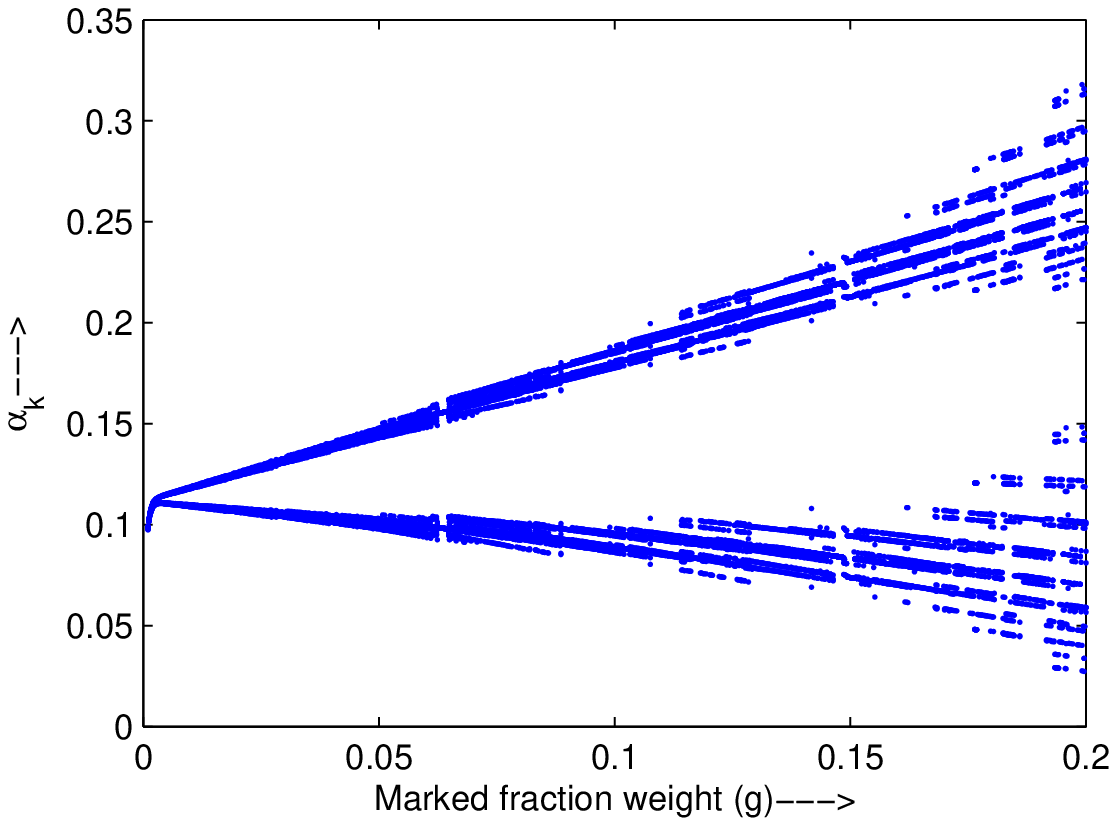}
      \caption {\scriptsize  The bifurcation diagram of $\alpha $ with $g$ as the bifurcation parameter and $K=20$}
\end{figure}

Here in this case we see stable operation of the system when the delay is small and it generates complicated dynamical behavior when the delay increases.  When the delay is large, the queue dynamics locks on several periodic attractors interleaved with ranges where cascades of bifurcations leading to higher periodic solutions may be detected. 
 \subsection{\normalsize Bifurcation parameter : Marking Threshold ($K$)}
We present the bifurcation diagram  with $K$ as the bifurcation parameter(Fig. 8) which varies from 5 to 25 through a step size of 0.01 ,and other parameters are given as. \\
$d=1 \mu s;\gamma=1;g=1/16. $\\
In this case we see similar behavior in the bifurcation diagram i. e the system alternating between stable periodic orbits and then locking on to high periodic orbits through period adding cascades as $K$ is varied.  

\subsection{Bifurcations in the congestion history variable $(\alpha)$}
We draw the bifurcation diagram of $\alpha_k$ with respect to the marked fraction weight ($g$) other parameters set as $\gamma=1$ and $d=100\mu s$ (Fig. 9).  $g$ varies from 0.001 to 0.2 through a step size of 0.0001. We see a novel type of bifurcation diagram which has a chaotic regime at low values of $g$ and shows high periodic orbits of $\alpha$ and larger amplitude of oscillation at higher values of $g$. This result is consistent with the relation defined in \eqref{alph}. The bifurcation diagram changes form as other parameters like delay($d$) ,$(\gamma )$ etc. are varied. 
\subsection{Time series plots and cobweb diagrams}
Here we show the actual time series evolution of the state variables for two different cases. One for a stable periodic regime and other for a chaotic regime. We also demonstrate the congestion control mechanism through \emph{cobweb diagrams} of the evolution of the window size. \\
case 1:
We take $d=30ns, \gamma=1, g=0.037$. The state space plots are given in Fig. 10 . As per the bifurcation diagrams the time series plots show a period 3 oscillatory behavior for all the variables. \\
case 2:
We take $d=30ns, \gamma=1, g=0.042$. The state space plots are given in Fig. 12. 
\begin{figure}[tphb]

\centering
\includegraphics[scale=0.45]{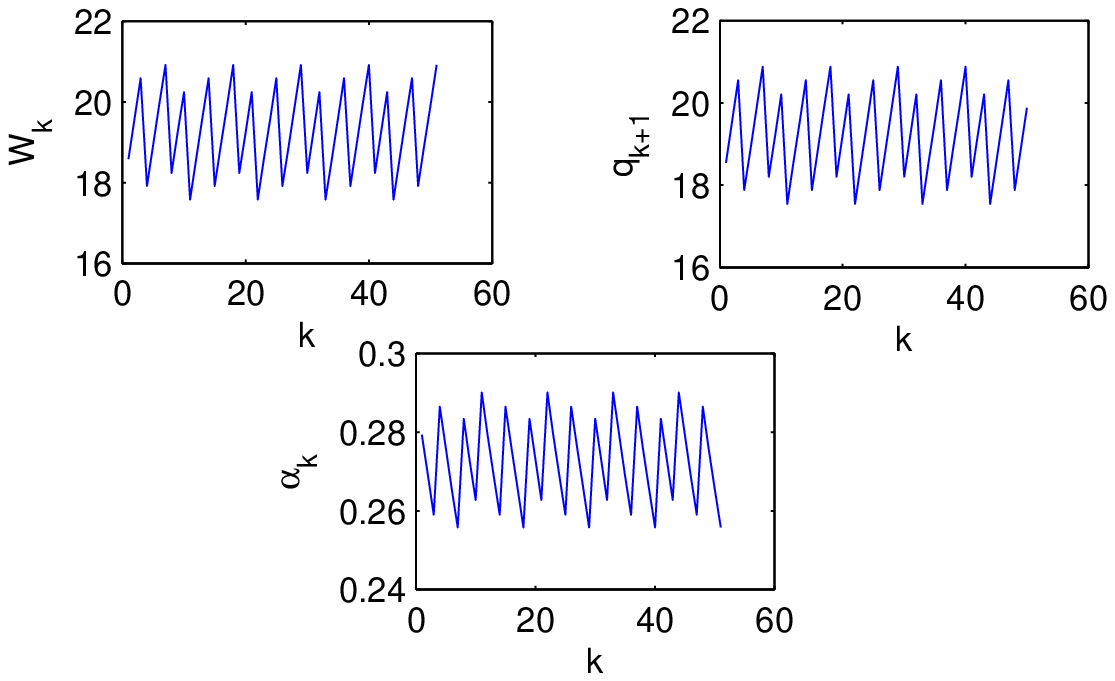}
\caption{The time series plots for case 1. }

\includegraphics[scale=0.45]{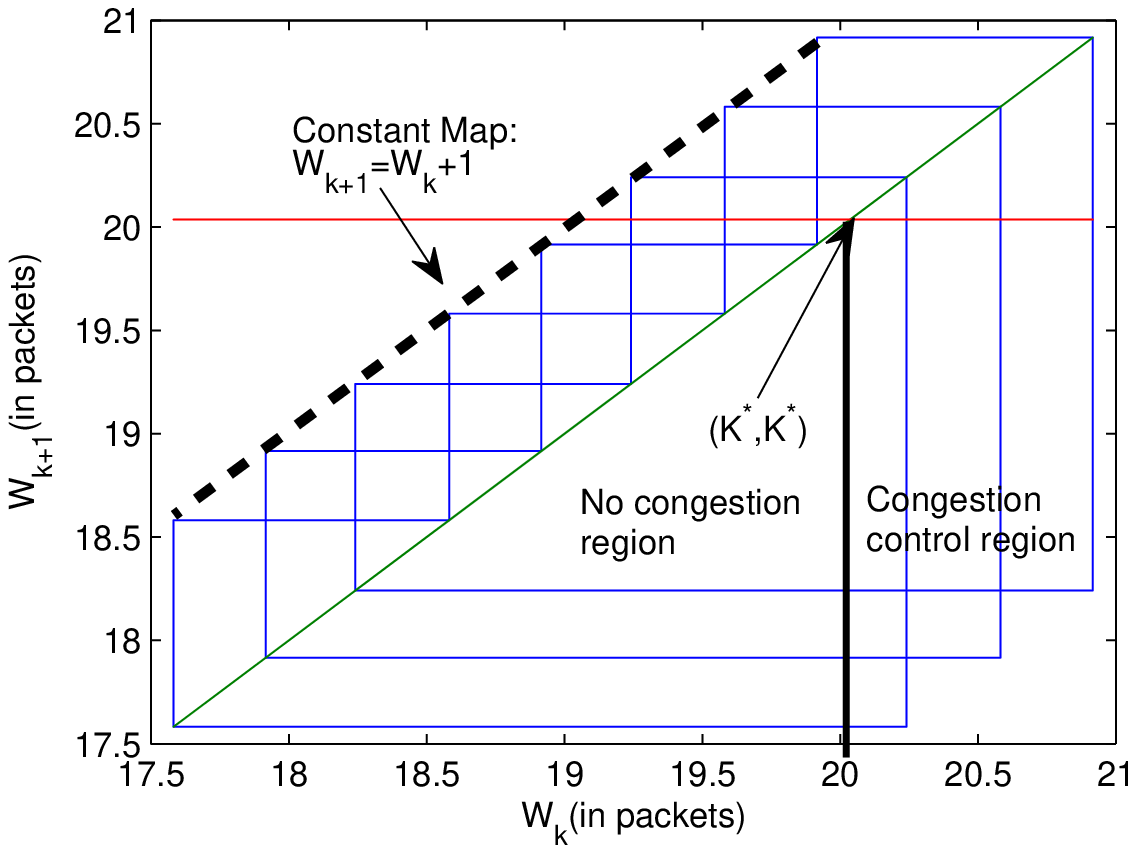}
\caption{Cobweb diagram for the evolution of $W$when $g=0. 037$. }
\centering
\includegraphics[scale=0.45]{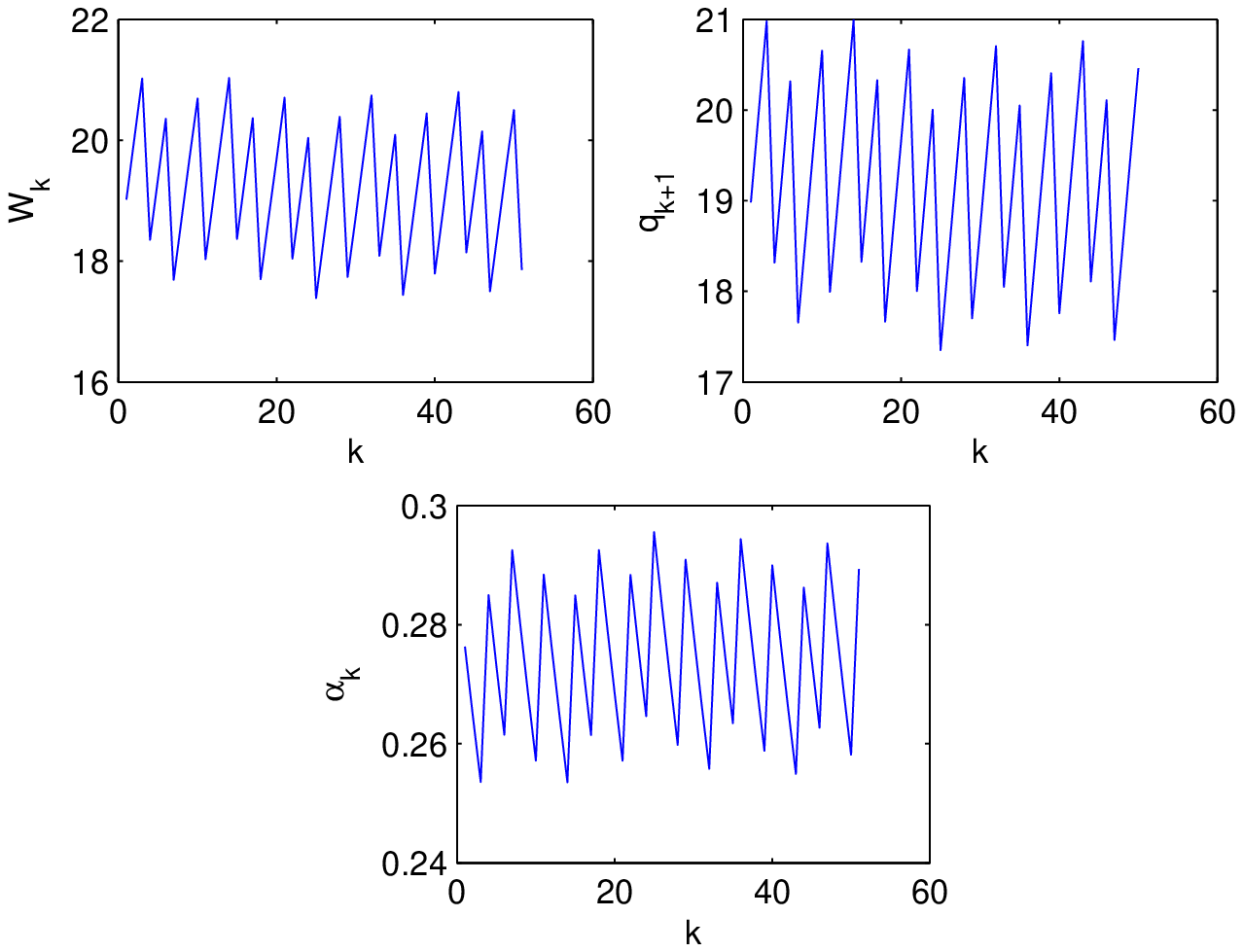}
\caption{The time series plots for case:2}
\centering
\includegraphics[scale=0.45]{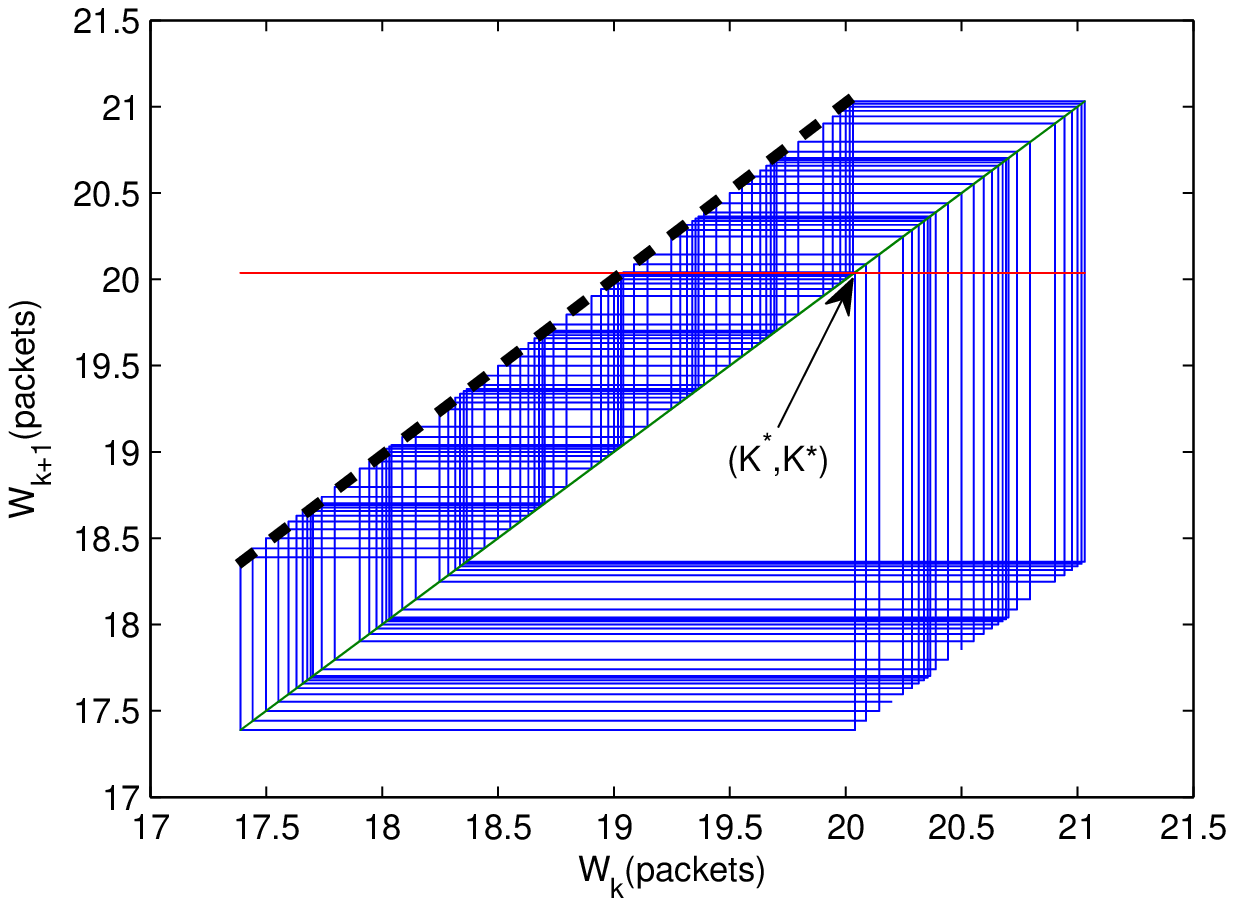}

\caption{Cobweb diagram for the evolution of $W$ when $g=0. 042$ showing chaotic dynamics }

\end{figure}

The cobweb diagram is widely employed to study the qualitative behavior of a map and it is possible to infer the long term status of an initial condition under repeated application of a map using it.  A cobweb diagram demonstrates graphically the successive mappings of the chosen state variable,  in our case $ W_k$. To plot the cobweb diagram we evolve the dynamical system starting from arbitrary initial condition and a given parameter setup long enough for it (if at all) to reach steady state behavior. Taking, the last n points of the steady state response of the chosen variable$(W_k)$, the successive mappings can be graphically shown by iteratively joining the  points($W_k,W_{k+1}$),($W_{k+1},W_{k+1}$),($W_{k+1},W_{k+2}$), where {$k=1,2,\ldots,(n-2)$}.  The number of lobes in the cobweb diagram communicates the period of steady state oscillation.  The cobweb diagram for \emph{case:1} is given in Fig. 11.  It shows 3 lobes thus implying a steady state period 3 behavior under the given circumstances.  The cobweb diagram for\emph{case:2} is given in Fig. 13.  It shows multiple lobes thus implying a very high periodicity and even possibly chaotic dynamics of the evolution of the window size.

\section{Generic nature of dynamical behavior and conclusion}
Nonlinear phenomena has been observed before in computer and telecommunication networks and our current work shows that even the most advanced next generation network protocols continue to display fully developed chaos which is in stark contrast with claims of global stable multi-frequency limit cycle behavior of the DCTCP system \cite{bal1}.  To make it mathematically precise,  almost all network protocols like TCP and their dynamics can be understood in terms of the iterations of the two-dimensional, piecewise nonlinear, discontinuous map  which is formulated, analyzed and simulated in this work.  The nature of the bifurcations reported here has a resemblance to the\emph{ corner bifurcations} reported in \cite{g2} which is observed in impact oscillators and it also shows similarity with chaotic behavior observed  earlier in current controlled boost converters\cite{s1}.  Future work involves demonstrating the DTCP chaos in dynamics\cite{t1}/ns-3\cite{t2}, in hardware and development of a new class of protocols for rock-stable and super-efficient operation over wildly fluctuating parameter regions.

\end{document}